\documentstyle[twocolumn,aps,epsf]{revtex}

\begin{document}

\draft

\title{ From Single to Multiple-Photon Decoherence in an Atom
Interferometer }

\author{ David A. Kokorowski, Alexander D. Cronin, Tony D. Roberts, and
David E. Pritchard }

\address{ Massachusetts Institute of Technology, Cambridge, MA 02139 }

\date{\today}

\maketitle
\newcommand{\betat}{\beta_{\rm total}}

\newcommand{\s}{\ ^{3}S_{\frac{1}{2}} }
\newcommand{\p}{\ ^{3}P_{\frac{3}{2}} }

\begin{abstract}
We measure the decoherence of a spatially separated atomic
superposition due to spontaneous photon scattering.  We observe a
qualitative change in decoherence versus separation as the number of
scattered photons increases, and verify quantitatively the decoherence
rate constant in the many-photon limit.  Our results illustrate an
evolution of decoherence consistent with general models developed for
a broad class of decoherence phenomenon.

\end{abstract}


% 39.20.+q   Atom interferometry techniques

% 03.75.-bÊÊÊMatter waves (for atom interferometry techniques, see 39.20Ñin atomic and molecular physics) 

% 03.75.BeÊÊÊAtom and neutron optics 

% 03.75.DgÊÊÊAtom and neutron interferometry

% 03.65.BzÊÊ Foundations, theory of measurement, miscellaneous theories (including AharonovÐBohm effect, Bell inequalities, Berry's phase) 
 
% 03.67.LxÊÊ Quantum computation

% 42.50.VkÊÊ Mechanical effects of light on atoms, molecules, electrons, and ions (see also 32.80.P and 33.80.P Optical cooling and trapping of atoms and molecules, respectively) 

% 32.80.LgÊÊ Mechanical effects of light on atoms, molecules, and ions

\narrowtext

Decoherence is the result of entanglement between a quantum system and
an unobserved environment, and manifests as the reduction of coherent
superpositions into incoherent mixtures.  This reduction occurs more
quickly as the number of particles comprising a quantum system
increases, establishing decoherence as a fundamental limit to
large-scale quantum computation\cite{UNR95} and
communication\cite{QCBOOK_COMM}.  Progress in these fields therefore
relies upon understanding and correcting for decoherence effects.  On
a macroscopic scale, decoherence is unavoidable and explains the
emergence of classical behavior in a world governed by quantum
mechanical laws.

Theoretical treatments of decoherence provide a description for the
evolution of a system's density matrix under the influence of a
specific environment.  For spatial decoherence, various environments
including a thermal bath of harmonic oscillators~\cite{CAL83}, a
scalar field~\cite{UNZ89}, and an isotropic distribution of
scatterers~\cite{JOZ85,GAF90} have been studied.  In the
high-temperature or many scatterer limit, these models all yield a
diffusion-like master equation for the system's spatial density
matrix, $\rho(x,x^{\prime})$:
\begin{equation}
	\frac{d\rho}{dt} = -\frac{i}{\hbar}\left[H,\rho\right]-
	D^{2}\,|x-x^{\prime}|^{2}\,\rho,
	\label{eqn:masterequation}
\end{equation}
where $H$ is the Hamiltonian for the isolated system and $D$, the
diffusion constant, depends on the details of the system-environment
coupling.  Assuming negligible internal dynamics, this equation
predicts an exponential reduction in coherence with time and with
separation squared\cite{WALLSNOTE}:
\begin{equation}
	\label{eqn:solvemaster}
	\rho(x,x^{\prime},t) = 
	e^{-D^{2}\,|x-x^{\prime}|^{2}\,t}\,\rho(x,x^{\prime},0).
\end{equation}
Similar decoherence behavior arises and has been studied in the
context of an atom interacting with a high-Q cavity\cite{BHD96}, and
trapped ions interacting with a fluctuating electric
field\cite{MKT00}.

To investigate the distinct case of decoherence due to scattering
processes, we have studied the loss of spatial coherence of atoms
within an atom interferometer due to spontaneous scattering of
photons.  In the many photon limit, this represents a simple case of
the general models above; we observe coherence loss consistent with
Eq.\ (\ref{eqn:solvemaster}) and are able to derive the decay constant
from first principles.  The few photon limit is of a qualitatively
different character, and we have followed the smooth transition
between these two regimes.

\begin{figure}
	\epsfxsize=3.25in
	\epsffile{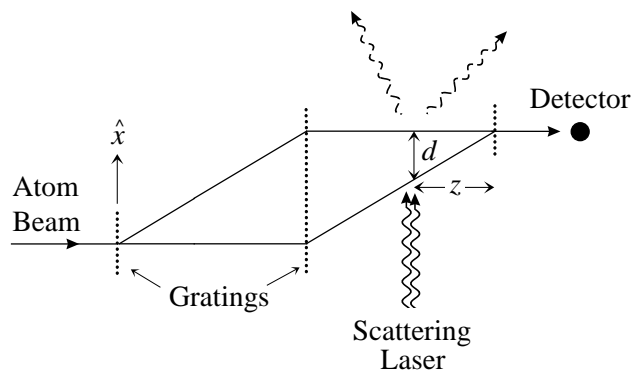}
	
	\caption{A schematic of our apparatus: a Mach-Zehnder
	interferometer comprised of three, evenly spaced, transmission
	gratings.  Within the interferometer, sodium atoms
	continuously absorb and spontaneously emit photons from a
	variable intensity laser beam.  Decoherence due to spontaneous
	emission results in reduced contrast interference fringes.}

	\label{fig:schematic}
\end{figure}

The atom interferometer\cite{AAMOP} is realized by passing a
collimated, supersonic beam of Na atoms (\mbox{velocity $\approx 3000$
m/s} using a He carrier gas) through three diffraction gratings
arranged in the Mach-Zehnder geometry (Fig.\ \ref{fig:schematic}). 
Prior to the first grating, the atoms are collimated and optically
pumped into the \mbox{$\s|F=2,m_{f}=+2\rangle$} ground state.  Two
paths through the interferometer, separated by up to $20 \mu$m,
overlap at the position of the third grating, forming a spatial
interference pattern.  This pattern is masked by the third grating and
the total transmitted flux is detected using a $50 \mu$m hot wire. 
The interference pattern is measured as an oscillating atomic flux
versus grating position.  Because the contrast of the interference
pattern is proportional to the coherence between the two paths,
reduction in contrast is direct evidence of coherence loss.

The effective decohering environment consists of photons from a laser
beam directed along the $\hat{x}$ axis which intersects both
interfering paths.  The circularly polarized laser light is tuned to
the \mbox{$\s|2,+2\rangle\rightarrow\p|3,+3\rangle$} transition with
wavelength \mbox{$\lambda=2\pi/k_{0} = 590$ nm}.  Because the atoms
are dipole forbidden from decaying to any state other than
$\s|2,+2\rangle$, they can continuously scatter photons without
falling out of resonance (the natural linewidth is $\sim 200$ photon
recoils wide).

At the intersection of the atomic beam and scattering laser, each
atom's transverse wavefunction is peaked at two positions which we
label $x$ and $x+d$.  If a photon, initially in momentum state
$|k_{0}\rangle$, scatters from this atom, the two become entangled:
\begin{eqnarray}
	\label{eqn:idealmeas}
	\big|\psi\big\rangle_{i} & = &\Big(|x\rangle +  |x+d\rangle\Big) 
	\otimes|k_{0}\rangle \stackrel{\mathrm{scat.}}{\longrightarrow} \nonumber \\
	&& \big|x\big\rangle\otimes\big|\phi_{x}\big\rangle + 
	\big|x+d\big\rangle\otimes e^{ik_{0}d}\big|\phi_{x+d}\big\rangle,
\end{eqnarray}
where $|\phi_{x}\rangle$ is the wavefunction of a photon spontaneously
emitted from position $x$ and the factor $e^{ik_{0}d}$ accounts for
the difference in spatial phase of the initial photon at the two
positions.  Generalizing the entangled wavefunction in Eq.\
\ref{eqn:idealmeas} to a density matrix and tracing over a basis of
scattered photon states, the net effect of scattering on the atom's
spatial density matrix is:
\begin{equation}
	\label{eqn:reducedrho}
	\rho(x,x+d) \stackrel{\mathrm{scat.}}{\longrightarrow} \rho(x,x+d)\,\beta(d),
\end{equation}
where $\beta(d)$ is known as the decoherence function and has the
properties $|\beta(d)|\leq 1$ and $\beta(0)=1$.  The decoherence
function thus defined is equal to the inner product of the two final
photon states, which are identical apart from an overall translation:
\begin{eqnarray}
	\label{eqn:beta}
	\beta(d) & = & e^{ik_{0}d}\,\langle\phi_{x}|\phi_{x+d}\rangle =
	e^{ik_{0}d}\langle\phi_{x}|e^{-i\,\hat{k}_{x}\,d}|\phi_{x}\rangle
	\nonumber \\
	& = & \int\,d\Delta\!k\,P(\Delta\!k)\,e^{-i\Delta\!kd},
\end{eqnarray}
where the operator $\hat{k}_{x}$ is the generator of photon
translations along the $\hat{x}$ axis.  The resulting decoherence
function is the Fourier transform of a probability distribution
$P(\Delta\!k)$, with $\Delta\!k=k_{x}-k_{0}$ being the change in
momentum of the photon along the $\hat{x}$ axis.

\begin{figure}
   \epsfxsize=3.25in
   \epsffile{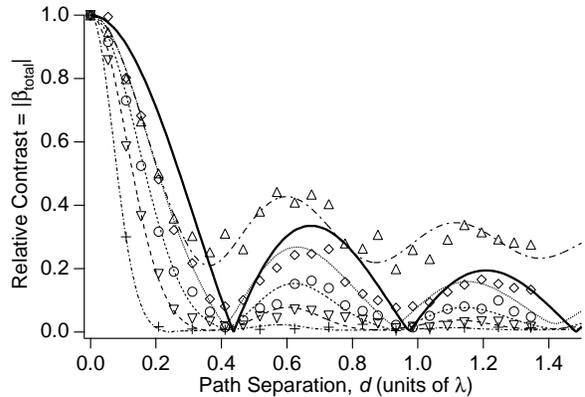}
   
   \caption{The total decoherence function, $|\betat|$, measured as
   the normalized contrast after spontaneous photon scattering.  The
   solid line is the single photon decoherence function.  Also
   displayed are the best fits from which we determine $\bar{n}=$ 0.9
   ($\triangle$), 1.4 ($\diamond$), 1.8 ($\circ$), 2.6
   ($\bigtriangledown$), and 8.2 ($+$).}
 
   \label{fig:nongaussian}
\end{figure}

Previous experiments\cite{PSK94,CHL95} have measured the decoherence
function for an atom which spontaneously scatters a single photon. 
The theoretical prediction which these experiments confirm is
displayed as the solid line in Fig.\ \ref{fig:nongaussian}.  Beneath
an overall decay in coherence with distance, periodic coherence
revivals are observed.  This shape follows directly from the Fourier
transform of the dipole radiation pattern for spontaneous emission. 
It has also been explained in terms of the ability of a single photon
to provide which-path information\cite{CHL95}: the contrast drops to
zero when the path separation is approximately equal to the resolving
power of an ideal Heisenberg microscope $d\approx\lambda/2$, with
revivals resulting from path ambiguity due to diffraction structure in
the image.   

If several photons are scattered, and if successive scattering events
are independent, the total decoherence function includes one factor of
$\beta$ for each scattered photon:
\begin{equation}
	\label{eqn:betat}
	\betat(d)=\sum_{n=0}^{\infty}P(n)\beta^{n}(d).
\end{equation}
In our experiment, the total number of photons scattered by an
individual atom is intrinsically uncertain, but is described by the
distribution $P(n)$ which can be measured or calculated.  The sum in
Eq.\ (\ref{eqn:betat}) is a trace over this additional degree of
freedom of the environment.

Figure\ \ref{fig:nongaussian} shows measurements of the decoherence
function for laser intensities corresponding to an average number of
scattered photons, $\bar{n}$, ranging from $\sim\!1$ to $\sim\!8$.  At
each intensity, a reference contrast and phase was measured, with the
scattering laser positioned such that the interfering paths were
completely overlapped ($d=0$).  We then adjusted the longitudinal
position of the scattering laser, $z$, to select specific path
separations in the range $0 < d < 1.4\lambda$ at which to measure the
decoherence function (see Fig.\ \ref{fig:schematic}).  For each path
separation, the ratio of the measured atom interference contrast to
the reference contrast yields the magnitude of the decoherence
function, $|\beta_{total}(d)|$.  The difference between the measured
atom interference phase and the reference phase yields the phase of
the decoherence function.

We fit the data using Eq.\ (\ref{eqn:betat}) and taking \mbox{$P(n)
\simeq {\rm
exp}\left(-\frac{1}{2}(n-\bar{n})^{2}/\sigma_{n}^{2}\right)$}.  This
form was chosen as a good approximation to Monte-Carlo Wavefunction
calculations of $P(n)$ for our laser parameters.  From the best fit
curves displayed in Fig.\ \ref{fig:nongaussian}, values were extracted
for $\bar{n}$ and $\sigma_{n}$ which were consistent with, and more
accurate than, independent measurements of $P(n)$ based on the
deflection and broadening of the atomic beam with the scattering laser
blocked versus unblocked.

In the regime $d\gg\lambda$, a single scattered photon suffices to
completely destroy the coherence between paths.  Thus, the non-zero
asymptotic value (for $\bar{n}=0.9$ in Fig.\ \ref{fig:nongaussian}) of
the decoherence function at large path separation is equal to the
fraction of atoms which scatter zero photons (i.e. decoherence is
proportional to the atom-photon interaction cross-section).  This
phenomenon is a simple example of saturation of
decoherence\cite{GAF90,APZ97}: the loss of coherence becomes
independent of path separation at a characteristic length scale of the
environment.  A recent experiment by Cheng and Raymer\cite{CHR99},
involving loss of optical coherence due to a disordered collection of
polystyrene microspheres, has features similar to our own: contrast
loss was observed to saturate when the path separation reached roughly
the diameter of the microspheres, and the asymptotic contrast was
proportional to the microsphere-light scattering cross section.

As the average number of scattered photons increases, the overall
amount of decoherence increases, and the contrast revivals disappear. 
This behavior can be formalized as the Fourier transform of the total
momentum distribution of all scattered photons:
\begin{equation}
	\label{eqn:betan}
	\beta^{n}(d) = \int\,d\Delta\!K\,P(\Delta\!K)\,e^{i\Delta\!K d},
\end{equation}
where $\Delta\!K = \sum_{i=1}^{n}\Delta\!k_{i}$.  As
$n\rightarrow\infty$, the central limit theorem predicts that
$P(\Delta\!K)$ will tend towards a Gaussian with mean $nk_{0}$ and
variance $n\sigma_{k}^{2}$ (where $\sigma_{k}=\frac{2}{5}k_{0}$ is the
rms transverse momentum of an emitted photon).  In the case of
spontaneous emission, $P(\Delta\!K)$ is approximately Gaussian for $n
> 3$ and the decoherence function reduces to:
\begin{eqnarray}
	\label{eqn:betaneval}
	\beta^{n}(d) & = &
	\int\,d\Delta\!K\,\left[e^{-\frac{1}{2}(\Delta\!K-nk_{0})^2/n\sigma_{k}^{2}} 
	\right]\,e^{i\Delta\!Kd} \\
	& = & e^{-\frac{1}{2}n\sigma_{k}^{2}d^{2}}e^{-ink_{0}d}. \nonumber
\end{eqnarray}
Inserting this expression into Eq.\ (\ref{eqn:betat}) and taking 
$d/\lambda \ll 1$, we find:
\begin{equation}
	\label{eqn:betatgaussian}
	\lim_{\bar{n}\rightarrow\infty}\betat(d) = 
	e^{-\frac{1}{2}\kappa^{2}d^{2}}e^{-i\bar{n}k_{0}d},
\end{equation}
where
\begin{equation}
	\label{eqn:kappa}
	\kappa^{2}=\bar{n}\sigma^{2}_{k}+\sigma^{2}_{n}k_{0}^{2}
\end{equation}
is the variance of the total momentum transferred to the atom from the
scattered photons.  The first term in Eq.\ (\ref{eqn:kappa}) comes from
the trace over modes available to the spontaneously emitted photon,
while the second is related to the uncertainty in number of absorbed
photons combined with the fixed phase $k_{0}d$ imparted by each.

If $\sigma_{n} = \sqrt{\bar{n}}$ (i.e. Poissonian statistics), Eq.\
(\ref{eqn:betatgaussian}) predicts an exponential decay in contrast
with number of scattered photons ($\kappa^{2}\propto\bar{n}$).  If in
addition the scattering rate, $\Gamma$, is constant, then
$\bar{n}=\Gamma t$ and the decoherence has exactly the exponential
form derived from a master equation like Eq.\
(\ref{eqn:masterequation}).

\begin{figure}
	\epsfxsize=3.25in
	\epsffile{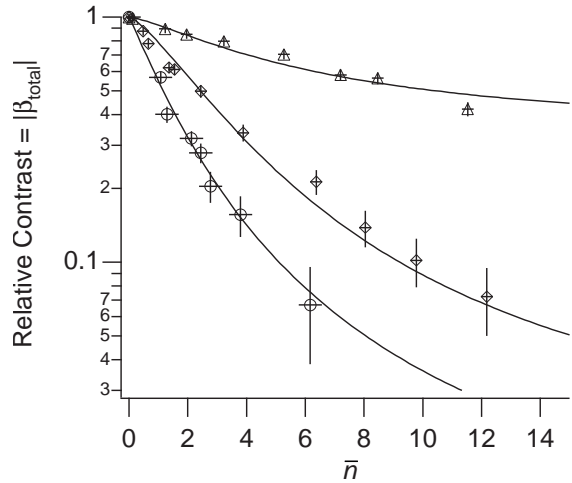}
	
	\caption{Loss of interfering contrast as a function of mean number
	of photons spontaneously scattered by atoms within the
	interferometer.  Each curve represents a different path
	separation: $d/\lambda = 0.06 (\triangle), 0.13 (\diamond),$ and $
	0.16 (\circ)$.}

	\label{fig:cvn}
\end{figure}

We have measured this exponential reduction of spatial coherence by
varying the average number of scattered photons, leaving the path
separation fixed (Fig.\ \ref{fig:cvn}).  Theory curves (solid lines)
are based on Eq.\ (\ref{eqn:betatgaussian}) with $\sigma_{n}$
determined from the broadening of the atomic beam due to the momentum
of the scattered photons.  The product of the two remaining free
parameters, $\bar{n}d$, was obtained from the measured phase of the
decoherence function.

The data follow a nearly exponential decay with $\bar{n}$.  The upward
trend at large $\bar{n}$ is a result of the finite size of our
hot-wire: the trace over final photon states (Eq.~\ref{eqn:betaneval})
must be restricted to those states which allow the atom to reach the
detector.  As a result $\kappa$ in Eq.\ (\ref{eqn:betatgaussian}) is
replaced with $\kappa^{\prime}$ where
$1/\kappa^{\prime\,2}=1/\kappa^{2}+1/\kappa_{d}^{2}$ and $\kappa_{d} =
3.3(1)k_{0}$ is our detector's effective momentum acceptance.

In the previous single-photon experiment of Chapman et~al.\cite{CHL95}
lost coherence was similarly ``recovered'' by positioning a hot-wire
detector to count only atoms which had scattered photons into a small
range of momentum states.  This scheme required\cite{WHC97} that the
atomic beam width, $\sigma_{x}$, be greater than the path separation,
$d$, so that the two interfering paths partially overlapped at the
point of scattering, and a scattered photon could not have provided
complete which-path information, even if $d\gg\lambda$.  The condition
$\sigma_{x}<d$ need not be satisfied to demonstrate the features of
decoherence in the current experiment, however.  Even when it is in
principle possible to recover some coherence by measuring the
environment, if no such attempt is made then the predicted loss of
contrast is independent of $\sigma_{x}$.

In the many photon limit, the decoherence function we have derived
agrees with the solution to the master equation presented in the
introduction.  Comparing Eqs.\ (\ref{eqn:solvemaster}) and
(\ref{eqn:betatgaussian}), taking into account the time varying
intensity profile, $I(t)$, of the scattering light as experienced by
atoms in the beam, we identify: $\kappa^{2}=D^{2}\tau$ where $\tau$ is
the amount of time needed to scatter $\bar{n}$ photons
($\bar{n}=\int_{0}^{\tau}\Gamma(I(t))\,dt$).  Because the atom-photon
scattering interaction is well defined, and our decohering environment
well controlled, we can accurately calculate the constant $\kappa$
(equivalently $D$) for any laser parameters.

\begin{figure}
   \epsfxsize=3.25in 
   \epsffile{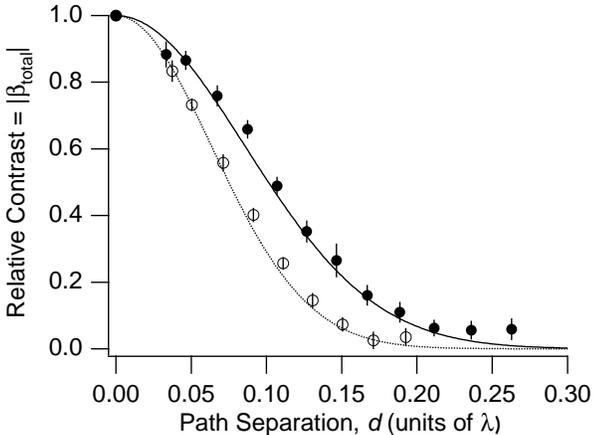} 

   \caption{Loss of contrast in the many-photon regime.  Overlayed are
   theory curves generated from Eq.\ (\ref{eqn:betatgaussian}) using
   parameters ($\bullet$) $\bar{n} = 4.8(2)$, $\sigma_{n}=1.8(1)$ and
   ($\circ$) $\bar{n} = 8.1(3)$, $\sigma_{n}=3.5(1)$ determined from
   independent beam deflection measurements.}

   \label{fig:gaussian}
\end{figure}

Displayed in Figure\ \ref{fig:gaussian} are data which demonstrate
Gaussian reduction in contrast as a function of path separation for
two different laser intensities.  As before, we independently
determined $\bar{n}$ and $\sigma_{n}$ for each intensity, and from
these values along with $\kappa_{d}$ we calculate
$\kappa^{\prime}=2.5(1)k_{0}$ for the higher laser intensity and
$\kappa^{\prime}=1.8(1)k_{0}$ for the lower.  Fitting the contrast
data to Eq.\ (\ref{eqn:betatgaussian}) yields
$\kappa^{\prime}=2.39(5)k_{0}$ and $\kappa^{\prime}=1.71(5)k_{0}$,
within error of the calculated values.

Our system exhibits what have been referred to as the
``naive''\cite{APZ97} generalizations of decoherence phenomenon:
exponential loss of contrast with path separation squared and with
number of scattered particles.  The similarity of Eq.\
(\ref{eqn:masterequation}) to a diffusion equation\cite{ZUR91},
invites identification of this type of decoherence with phase
diffusion or a random phase walk.  To make the identification
explicit, we use the identity $|\phi_{x+d}\big\rangle =
e^{-i\hat{k}_{x}d}|\phi_{x}\big\rangle$ to rewrite Eq.\
(\ref{eqn:idealmeas}) as:
\begin{eqnarray}
	\label{eqn:meastodiff}
	&&\big|\psi\big\rangle_{i} \stackrel{\mathrm{scat.}}{\longrightarrow} 
	\big|x\big\rangle\otimes\big|\phi_{x}\big\rangle + 
	\big|x+d\big\rangle\otimes e^{ik_{0}d}e^{-i\hat{k}_{x}d}\big|\phi_{x}\big\rangle 
	= \nonumber \\
	& &\int\,d\vec{k}\,\left(|x\rangle + e^{-i(k_{x}-k_{0})d}|x+d\rangle\right)
	\otimes|\vec{k}\rangle\langle \vec{k}|\phi_{x}\rangle
\end{eqnarray}
In this expression for the entangled atom-photon wavefunction, a
photon state $|\vec{k}\rangle$ corresponds to an atomic superposition
state with the phase between the two components shifted by an amount
$\Delta\!  \phi = (k_{x}-k_{0})d$.  Correlating interference
data with measurements of each scattered photon momentum (effectively
a randomly sampled element of the distribution $P(k)$) would allow
complete recovery of lost contrast.  In the absence of such
post-processing, however, the phase of each atom's interference
fringes will vary randomly, and their sum, the measured interference
pattern, will have reduced contrast.  The phase diffusion and
(previously discussed) which-path pictures are equally valid when the
experimenter does not measure the scattered photons\cite{SAI90}.

In conclusion, we have studied the decoherence of a spatial
superposition due to photon scattering.  Our data confirm theoretical
predictions, and in the many-photon limit exhibit features of
decoherence which are quite general.  We have observed the exponential
coherence loss with time and path separation squared characteristic of
this general behavior, and we have for the first time predicted and
experimentally verified the decoherence rate constant $\kappa$.  The
particular model we have explored is not only the most relevant for
macroscopic systems but also applies generally to situations in which
decoherence arises slowly though a series of independent, mildly
decohering interactions, a situation of interest for decoherence
avoidance or correction protocols.

This work was supported by Army Research Office contracts
DAAG55-98-1-0429 and DAA55-97-1-0236, Office of Naval Research
contract N00014-96-1-0432, and National Science Foundation grant
PHY-9877041.

\end{document}